# Prominent ethanol sensing with $Cr_2O_3$ nanoparticle-decorated ZnS nanorods sensors

**Gun-Joo Sun, Hyejoon Kheel, Taegyung Ko and Chongmu Lee**

*[1]Department of Materials Science and Engineering, Inha University, 253 Yonghyun-dong, Nam-gu, Incheon 402-751, Republic of Korea*

**Hyoun Woo Kim**

*[2]Department of Materials Science and Engineering, Hanyang University, 253 Haengdang-dong, Seongdong-gu, Seoul 402-751, Republic of Korea*

ZnS nanorods and $Cr_2O_3$ nanoparticle-decorated ZnS nanorods were synthesized using facile hydrothermal techniques and their ethanol sensing properties were examined. X-ray diffraction and scanning electron microscopy revealed the good crystallinity and size uniformity of the ZnS nanorods. The $Cr_2O_3$ nanoparticle-decorated ZnS nanorod sensor showed stronger response to ethanol than the pristine ZnS nanorod sensor. The responses of the pristine and decorated nanorod sensors to 200 ppm of ethanol at 300°C were 2.9 and 13.8, respectively. Furthermore, under these conditions, the decorated nanorod sensor showed longer response time (23 s) and shorter recovery time (20 s) than those of the pristine one (19 and 35 s, respectively). Consequently, the total sensing time of the decorated nanorod sensor (42 s) was shorter than that of the pristine one (55 s). The decorated nanorod sensor showed excellent selectivity to ethanol over other volatile organic compound gases including acetone, methanol, benzene, toluene, whereas the pristine one failed to show selectivity to ethanol over acetone. The improved sensing performance of the decorated nanorod sensor is attributed to the modulation of the conduction channel width and the potential barrier height at the ZnS-$Cr_2O_3$ interface accompanying the adsorption and desorption of ethanol gas as well as the greater surface–to-volume ratio of the decorated nanorods than the pristine one due to the existence of the ZnS-$Cr_2O_3$ interface.

Email: cmlee@inha.ac.kr

Fax: +82-32-862-5546

## I. INTRODUCTION

The remarkable fundamental properties of zinc sulfide (ZnS) has made them have diverse applications such as light-emitting diodes, lasers, flat panel displays, infrared windows, sensors, and biodevices, etc [1]. In particular, ZnS can be applied to the fabrication of ultraviolet (UV) light sensors and gas sensors. Over the past decade, the following ZnS nanostructure based gas-sensors have been reported: a ZnS single nanobelt sensor for $H_2$ sensing [1], ZnS nanobelt sensors for $H_2$ sensing [2], ZnS microsphere sensors for $O_2$ sensing [3], ZnS nanowire sensors for acetone and ethanol sensing [4], ZnS nanotube array sensors for humidity sensing [5]. However, reports on ZnS nanostructures-based gas sensors are very few compared to that of metal oxide semiconductor nanostructures-based gas sensors [1-5].

Metal oxide semiconductors are endowed with many good properties such as high sensitivity, fast sensing, low detection limits and function durability that the sensor materials should have. On the other hand, this sensing material has several shortcomings such as high operation temperature, poor selectivity and reliability. A range of techniques including doping, heterostructure formation and light activation have been studied to overcome these drawbacks. Of these techniques, the heterostructure formation technique was adopted to overcome the poor performance of the $SnO_2$1D nanostructure-based sensors in this study. Formation of heterostructures by creating interfaces between two dissimilar semiconducting materials can make the Fermi levels across the interface equal, i.e., in equilibrium, resulting in charge transfer and the formation of a interfacial depletion region. This will eventually lead to enhanced sensor performance. The enhanced sensing properties of these heterostructures might be attributed to many factors, including electronic effects such as band bending due to Fermi level equilibration, charge carrier separation, depletion layer manipulation and increased interfacial potential barrier energy, chemical effects such as decrease in activation energy, targeted catalytic activity and synergistic surface reactions and geometrical effects such as grain refinement, surface area enhancement, and increased gas accessibility [6]. Heterostructure formation is commonly achieved by either forming core-shell structures by coating nanostructures with a thin film or decorating nanostructures with dissimilar semiconductor nanoparticles. In this paper, we report the synthesis of $Cr_2O_3$ nanoparticle-decorated ZnS nanorods via a facile hydrothermal route and their enhanced sensing properties towards ethanol ($C_2H_5OH$) gas.

## II. EXPERIMENTAL PROCEDURE

### 1. Synthesis of pristine and $Cr_2O_3$ nanoparticle-decorated ZnS nanorods

The $Cr_2O_3$ nanoparticle-decorated ZnS nanorods were synthesized via a facile hydrothermal route: ZnS nanorods were synthesized using a hydrothermal method. First, Au-coated sapphire was used as a substrate for the synthesis of ZnS nanorods. Au was deposited on a silicon (100) substrate by direct current (dc) magnetron sputtering. A quartz tube was mounted horizontally inside a tube furnace. An alumina boat containing 99.99 % pure ZnS powders and silicon substrates were placed separately in the two-heating zone-tube furnace, where the ZnS powders were in the first heating zone and Si substrates in the second heating zone. The substrate temperatures of the first and second heating zones were set to 850 and 650$^{o}$C, respectively, with an ambient nitrogen gas pressure and a flow rate maintained at 1 Torr and 50 cm$^3$/min, respectively, throughout the synthesis process. The thermal evaporation process was carried out for 1 h and then the furnace was cooled to room temperature at 1 mTorr, after which the products were taken out.

On the other hand, 50-mM $Cr_2O_3$ precursor solution was prepared by dissolving chromium acetate monohydrate ($Cr(CH_3COO)_2 \cdot H_2O$) in distilled water. 50 ml of the $Cr_2O_3$ precursor solution and 10 ml of 28% $NH_4OH$ solution were mixed together. The mixed solution was then ultrasonicated for 30 min to form a uniform solution and then rotated using a centrifuge at 5,000 rpm for 2 min to precipitate the $Cr_2O_3$ powders. The precipitated powders were collected by removing the liquid leaving the powders behind. The collected powders were rinsed in a 1:1-solution of isopropyl alcohol (IPA) and distilled water to remove the impurities. The rinsing process was repeated five times. Subsequently, the $Cr_2O_3$ precursor solution was dropped onto the ZnS nanorods on a substrate and the substrate was rotated at 1,000 rpm for 30 s for $Cr_2O_3$ decoration. After the spin-coating process, the $Cr_2O_3$ decorated ZnS nanorod sample was dried at 150°C for 1 min and then annealed in air at 500°C for 1 h.

## 2. Materials characterization

The phase and crystallinity of the pristine and $Cr_2O_3$ nanoparticle-decorated ZnS nanorods were analyzed by XRD (Philips X'pert MRD) using Cu Kα radiation (1.5406Å). The data was collected over the 2θ range, 20–80°, with a step size of 0.05° 2θ at a scan speed of 0.05°/s. Assignment of the XRD peaks and identification of the crystalline phases were carried out by comparing the obtained data with the reference compounds in the JCPDS database. The morphology and particle size of the synthesized powders were examined by SEM (Hitachi S-4200) at an accelerating voltage of 5 kV.

## 3. Sensor Fabrication

For the sensing measurement, a $SiO_2$ film (~200 nm) was grown thermally on the single crystalline Si (100). In the meantime, the as-synthesized ZnS nanorods and $Cr_2O_3$ nanoparticle-decorated ZnS nanorods were dispersed in a 1:1 mixture of deionized water and isopropyl alcohol by ultrasonication. The $Cr_2O_3$ nanoparticle-decorated ZnS nanorod sensors were fabricated by pouring a few drops of nanorod-suspended ethanol onto the $SiO_2$-coated Si substrates equipped with a pair of interdigitated (IDE) Ni (~10 nm)/Au (~100 nm) electrodes with a 20 μm gap (Fig. 1).

## 4. Sensing tests

The electrical and gas sensing properties of the pristine and $Cr_2O_3$ nanoparticle-decorated ZnS nanorods were determined at different temperatures in a quartz tube inserted in an electrical furnace. During the tests, the nanorod gas sensors were placed in a sealed quartz tube with an electrical feed through. A predetermined amount of ethanol (>99.99 %) gas was injected into the testing tube through a microsyringe to obtain ethanol concentrations of 10, 20, 50, 100, and 200 ppm while the electrical resistance of the nanorods was monitored. The response was defined as $R_a/R_g$ where $R_g$ and $R_a$ are the electrical resistances of sensors in ethanol gas and air, respectively. The response time was defined as the time needed for the change in electrical resistance to reach 90% of the equilibrium value after injecting ethanol, and the recovery time was defined as the time needed for the sensor to return to 90 % of the original resistance in air after removing the ethanol gas.

# III. RESULTS AND DISCUSSIONS

## 1. Crystalline structure and morphology

The structure and chemical composition of pristine and $Cr_2O_3$ nanoparticle-decorated ZnS nanorods were examined by XRD immediately after sample preparation. As shown in Fig. 2, all the XRD peaks marked by the red circle were consistent with the standard value of the $Cr_2O_3$ phase (JCPDS No. 84-1616). The peaks marked by the black circle were attributed to the formation of ZnS phase (JCPDS No. 89-2942). The fact that no distinct peaks existed except the patterns of $Cr_2O_3$ and ZnS indicates that no other phases, such as had formed, highlighting the purity of the final products. The (200) plane was chosen to calculate the crystallite size of the pristine and $Cr_2O_3$ nanoparticle-decorated ZnS nanorods using the Scherrer formula [7].

$$D=K\lambda/\beta cos\theta \qquad (1)$$

where $D$ is the crystallite size in nm, $K$ is the shape factor (0.90), $\lambda$ is the wavelength of X-rays used (1.5406Å), $\beta$ is the full-width at half maximum in degrees and $\theta$ is the diffraction angle in degrees. The values obtained were 60 nm and 50 nm for the pristine and $Cr_2O_3$ nanoparticle-decorated ZnS nanorods, respectively.

Figures 3(a) and (b) show SEM images of pristine and $Cr_2O_3$ nanoparticle-decorated ZnS nanorods, respectively. The pristine ZnS nanorods were 50-100 nm in diameter and up to a few tens of micrometers in length. The entire surfaces of the ZnS nanorods in the decorated nanorods are completely covered with many lenticular shaped-$Cr_2O_3$ nanoparticles with long radii of 30-40 nm and small radii of 10 - 20 nm (Fig. 3(b)). The surface–to-volume ratio of the decorated nanorods must be much higher than that of the pristine nanorods.

## 2. Gas-sensing properties

### 2.1. Optimal working temperature

The sensitivity of gas sensors is strongly influenced by the operating temperature. Parallel experiments were carried out over the temperature range from 200 to 400°C to determine the optimal operating temperature of the sensors. Figure 4 shows the relationship between the response of the pristine and $Cr_2O_3$ nanoparticle-decorated ZnS nanorod sensors to 200 ppm of ethanol in a temperature range of 200-400°C. Both the pristine and $Cr_2O_3$ nanoparticle-decorated ZnS nanorod sensors showed a maximum response at 300°C, suggesting that 300°C is the optimal operating temperature for both sensors. The temperature dependence of the sensor response is generally controlled by two parameters: the reaction rate between the adsorbed oxygen ions with ethanol molecules, and the electron density of the sensor. The reaction rate coefficient and electron density increases exponentially with increasing temperature. On the other hand, the sensor response is proportional to the reaction rate coefficient and inversely proportional to the electron density. These two parameters compete with each other and result in a maximum sensor response at the optimal operating temperature [8].

### 2.2. Sensor response with ethanol gas concentration

Figures 5(a) and (b) present the gas response transients of the pristine and $Cr_2O_3$ nanoparticle-decorated ZnS nanorod sensors towards 5, 10, 50, 100, and 200 ppm of ethanol gas at 300°C, respectively. For both sensors, the response was fully reversible and both sensors exhibited n-type behavior upon exposure to ethanol gas.

Figure 6 shows the calibration curves for the responses of the two sensors to different ethanol concentrations, where it clearly shows that the response of the $Cr_2O_3$ nanoparticle-decorated ZnS nanorod sensor to every ethanol concentration is stronger than that of the pristine one. As an example, the response of the $Cr_2O_3$ nanoparticle-decorated ZnS nanorod sensor to 200 ppm of ethanol is approximately 4.5 times stronger than the pristine one.

The Occupational Safety Health Administration (OSHA) established the maximum recommended exposure level of ethanol to be 1000 ppm [9], and the $Cr_2O_3$ nanoparticle-decorated ZnS nanorod sensor can easily detect this level of ethanol.

A relationship between the sensor response ($S=R_g/R_a$) and ethanol concentration can be expressed as an empirical equation:

$$S = R_g/R_a = A\,[C_{ethanol}]^b + 1 \quad (2)$$

where $A$, $b$ and $[C_{ethanol}]$ are a constant, an exponent and the ethanol concentration, respectively [10]. In fact, "$b$" is a charge parameter with an ideal value of 1 for $O^-$ and 0.5 for $O^{2-}$, which is derived from the surface interaction between the chemisorbed oxygen and target gas [11,12]. The response of the decorated nanorod sensor tended to increase more rapidly than that of the pristine one as the ethanol concentration increased, suggesting that the response of the former to ethanol would be much stronger than that of the latter at high ethanol concentrations.

**2.2. Response and recovery times**

Figures 7(a) and (b) present the response and recovery times of both sensors towards 200-ppm ethanol at 300°C, respectively. As shown in Figs. 7(a) and (b), the response and recovery times of both sensors became shorter with increasing ethanol concentration. The change in response time can be explained by the change in the saturation time and the mean residence period of the ethanol molecules on the sensor surface. When the ethanol concentration is low, it spends a relatively long time reacting with adsorbed oxygen species. With increasing the ethanol concentration, more ethanol molecules are available for the reaction with adsorbed oxygen, resulting in a decrease in response time. The change in recovery time with the ethanol concentration can be explained by the structure of the sensors and diffusion rate. When air is injected into the test chamber, oxygen molecules will diffuse to the surface of the sensors to react with the ethanol molecules. The complete desorption reaction of the inner surface takes more time than that on the outer surface, leading to a longer recovery time at higher ethanol concentrations, where ethanol is present near the inner surfaces of the sensors.

In the case of the response time, for all ethanol concentrations, the $Cr_2O_3$ nanoparticle-decorated ZnS nanorod sensor had a slightly longer response time, which is probably due to the far higher resistance of the $Cr_2O_3$ nanoparticle-decorated ZnS nanorod sensor at 300°C (Fig. 5(b)). This means that there are less adsorbed oxygen species on the surface of the decorated nanorod sensor. Therefore, after injecting the ethanol gas, they react quite slowly with the adsorbed oxygen species, leading to a longer response time. The shorter recovery time of the $Cr_2O_3$ nanoparticle-decorated ZnS nanorod sensor is probably due to the faster desorption of ethanol gas because of the lower potential barrier of ethanol desorption in the particular structure of the $Cr_2O_3$ nanoparticle-decorated ZnS nanorod.

Figure 8 shows that the response of the pristine and $Cr_2O_3$ nanoparticle-decorated ZnS nanorod sensors to ethanol gas at 300°C was stronger than those to the other volatile organic compound (VOC) gases, indicating that both sensors had excellent selectivity towards ethanol gas. It is not fully understood why these ZnS-based sensors have selectivity to ethanol at 300°C over other gases. The selectivity might be related to the different optimal operating temperatures of the sensor for different target gases. The response of a sensor would depend strongly on the type of gas at different temperatures because different gases have different activation energies for adsorption, desorption and reaction on the semiconductor surface [13]. For these ZnS-based sensors, 300°C may be an optimal operating temperature because the activation energy for the adsorption of ethanol is low at that temperature, whereas those for the adsorption of other gas species are relatively high at that temperature.

Up to now, metal oxide sensors for ethanol detection have been studied extensively. This is because ethanol is used widely in different industries and its detection in drunk drivers is important for social safety. Table 1 compares the ethanol sensing properties of the pristine and $Cr_2O_3$ nanoparticle-decorated ZnS nanorod sensor fabricated in this study with other 1D gas sensors reported in the literature [14-24]. The table shows that the response and response/recovery times of the $Cr_2O_3$ nanoparticle-decorated ZnS nanorod sensor are comparable to those of metal oxide semiconductor 1D nanostructured sensors.

**2.3. Gas-sensing mechanism**

Based on the above results, the $Cr_2O_3$ nanoparticle-decorated ZnS nanorod sensor showed a significantly improved sensing performance compared to the pristine ZnS sensor. For the pristine ZnS sensor, the gas sensing mechanism can be explained mainly in terms of modulation of the depletion layer accompanying the adsorption and desorption of gases. When the pristine ZnS sensor is exposed to air, oxygen molecules are adsorbed on the surfaces of the ZnS nanorods and are ionized to either $O^-$ or $O^{2-}$ by capturing free electrons from the conduction band of ZnS. This reduces the electron concentration, which then leads to the formation of an electron depletion layer. When ZnS is exposed to ethanol gas, the ethanol molecules react with the oxygen species ($O^-$, $O^{2-}$) adsorbed on the surfaces of the ZnS nanorods according to the following equations [25]:

$$C_2H_5OH\ (gas) \rightarrow C_2H_5OH\ (ads) \quad (1)$$

$$C_2H_5OH\ (ads) + 6O^-\ (ads) \rightarrow 2CO_2\ (gas) + 3H_2O\ (gas) + 6e^- \quad (2)$$

These reactions release the trapped electrons back to the conduction band of ZnS, which increases the free electron concentration, and ultimately decreases the resistance of the pristine ZnS sensor.

On the other hand, for $Cr_2O_3$ nanoparticle-decorated ZnS nanorod sensor, ZnS and $Cr_2O_3$ are n-type and p-type semiconductors, respectively, with different electron affinities (3.8 eV for ZnS [26], no data available for $Cr_2O_3$). Little is known about the electron affinity of $Cr_2O_3$, but the Fermi energy level of ZnS might be lower than that of $Cr_2O_3$ because ZnS and $Cr_2O_3$ are n- and p-type semiconductors, respectively, and the Fermi energy level of a n-type semiconductor is commonly higher than that of a p-type semiconductor. Therefore, the transfer of electrons will occur from the conduction band of $Cr_2O_3$ to that of ZnS to make the two Fermi energy levels ($E_F$) equal (Fig. 9(a)). This will result in the formation of an electron depletion layer and a potential barrier at the ZnS-$Cr_2O_3$ n-p junction interface, which will enhance the response of the $Cr_2O_3$ nanoparticle-decorated ZnS nanorod sensor further compared to the pristine one.

The enhanced ethanol gas sensing performance of the $Cr_2O_3$ nanoparticle-decorated ZnS nanorod sensor can be explained by modulation of the conduction channel width [50] and the potential barrier height at the ZnS-$Cr_2O_3$ interface [27,28]. Figure 9 presents schematic diagrams showing the depletion layer and potential barrier formed at the ZnS-$Cr_2O_3$ interface as well as the energy band diagrams of the ZnS-$Cr_2O_3$ binary system in air and ethanol gas. The width of the depletion layers formed near the ZnS-$Cr_2O_3$ interfaces in the $Cr_2O_3$ nanoparticle-decorated ZnS nanorods is larger than that formed on the surface regions in the pristine ZnS nanorods; ($\lambda_D$(ZnS)+$\lambda_D$($Cr_2O_3$)) for the decorated nanorods, $\lambda_D$($ZnS_2$) for the pristine ZnS nanorods, where $\lambda_D$ is the Debye length. $\lambda_D$(ZnS) = $10^{-5}$-$10^{-6}$ cm [29], Even though the data of $\lambda_D$($Cr_2O_3$) is not available at present, a large portion of the total volume of each $Cr_2O_3$ nanoparticle might be depleted of carriers in air. The larger depletion layer width in the decorated nanorods than that in the pristine nanorods leads to higher resistivity and a larger change in resistivity. In addition to the increased depletion layer width, the formation of a potential barrier at the ZnS-$Cr_2O_3$ interface due to electron trapping in the interface states should be considered when explaining the enhanced response of the decorated nanorods to ethanol gas. Upon exposure to ethanol gas, the potential barrier at the ZnS-$Cr_2O_3$ interface will decrease, whereas after stopping the ethanol gas supply, the potential barrier will increase upon exposure to air (Fig. 9). Hence, modulation of the potential barrier occurs concomitantly with the adsorption and desorption of gas molecules, which would increase the change in resistance, i.e., the response of the decorated nanorod sensor to ethanol gas.

In addition to the above two effects, the ZnS-$Cr_2O_3$ interfaces provide additional. Preferential adsorption sites and diffusion paths for oxygen and ethanol molecules [30], which might also contribute to the enhanced ethanol gas sensing properties of the decorated nanorod sensor. In other words, the enhanced response of the $Cr_2O_3$ nanoparticle-decorated ZnS nanorod sensor is partially attributed to the higher surface-to-volume ratio of the decorated nanorods than that of the pristine one because the ZnS-$Cr_2O_3$ interfaces also act as preferential adsorption sites like the outer surface of the nanorods.

## IV. CONCLUSION

ZnS nanorods and $Cr_2O_3$ nanoparticle-decorated ZnS nanorods were synthesized by hydrothermal techniques. The $Cr_2O_3$ nanoparticle-decorated ZnS nanorod sensor exhibited a significantly stronger response to ethanol than pristine one. The decorated nanorod sensor showed a lower working temperature than the other oxide semiconductor 1D nanostructured ethanol gas sensors. This was attributed to the larger modulation of the depletion layer width and modulation of the potential barrier height at the ZnS-$Cr_2O_3$ interfaces in the ZnS nanorods and $Cr_2O_3$ nanoparticle-decorated ZnS nanorods as well as to the crystallographic defects formed at the ZnS-$Cr_2O_3$ interfaces acting as preferential adsorption sites and diffusion paths for gases. Furthermore, the sensors also showed excellent selectivity to ethanol. The favorable gas sensing performance makes $Cr_2O_3$ nanoparticle-decorated ZnS nanorods particularly attractive as a promising practical sensor material.

## ACKNOWLEDGMENTS

This study was supported by the MSIP (Ministry of Science, ICT and Future Planning), Korea, under the C-ITRC (Convergence Information Technology Research Center) (IITP-2015-H8601-15-1003) supervised by the IITP (Institute for Information & communications Technology Promotion) and the Basic Science Research Program through the National Research Foundation of Korea (NRF) funded by the Ministry of Education (2010-0020163).

Table 1. Comparison of the response, response time and recovery time of the $Cr_2O_3$ nanoparticle-decorated ZnS nanorod sensor with those of other material 1D nanostructure sensors reported previously.

| Nanomaterials | Ethanol conc. (ppm) | Temp. (°C) | Response (%) | Response Time (sec) | Recovery Time (sec) | Ref. |
|---|---|---|---|---|---|---|
| ZnS/ nanorods | 200 | 300 | 290 | 19 | 35 | Present work |
| ZnS/$Cr_2O_3$ nanorods | 200 | 300 | 1,384 | 23 | 20 | Present work |
| $TiO_2$ nanotubes | 5,000 | 200 | 16 | - | - | [14] |
| $SnO_2$ nanorods | 300 | 300 | 3,140 | 1 | 1 | [15] |
| Ce-$SnO_2$ nanopowders | 200 | 250-450 | 18,500 | - | - | [16] |
| Pt-$SnO_2$ nanopowders | 100 | 150-350 | 4,000 | 12 | 360 | [17] |
| $SnO_2$-ZnO(0.05) composite nanopowders | 300 | 200-400 | 390,000 | 96-418 | 400-600 | [18] |
| ZnO-$SnO_2$(0.05) composite nanopowders | 300 | 200-400 | 120,000 | 96-418 | 400-600 | [18] |
| ZnO nanowires | 1,500 | 300 | 61 | - | - | [19] |
| $TiO_2$ nanobelts | 500 | 250 | 3,366 | 1-2 | 1-2 | [20] |
| Ag-$TiO_2$ nanobelt | 500 | 200 | 4,171 | 1-2 | 1-2 | [20] |
| $CoFe_2O_4$ nanopowders | 50 | 150 | 7190 | 50 | 60 | [21] |
| Co-ZnO nanorods | 50 | 350 | 987 | - | - | [22] |
| $In_2O_3$ nanowires | 100 | 370 | 200 | 10 | 20 | [23] |
| $In_2O_3$ nanorods | 50 | - | 795 | 5 | 10 | [24] |

## Figure Captions

Fig. 1. (a) Schematic diagram of the synthesis procedure for the $Cr_2O_3$ nanoparticle-decorated ZnS nanorods. (b) Schematic of the sensor structure.

Fig. 2. XRD patterns of the pristine and $Cr_2O_3$ nanoparticle-decorated ZnS nanorods

Fig. 3. SEM image (a) the pristine ZnS nanorod and (b)the $Cr_2O_3$ nanoparticle-decorated ZnS nanorods.

Fig. 4. Response of the pristine and $Cr_2O_3$ nanoparticle-decorated ZnS sensors to 200 ppm of ethanol at different temperatures.

Fig. 5. Gas response transients of (a) the pristine ZnS nanorod sensor and (b) $Cr_2O_3$ nanoparticle-decorated ZnS nanorod sensors towards 5, 10, 50, 100, and 200 ppm of ethanol gas at 300°C.

Fig. 6. Responses of the pristine and $Cr_2O_3$ nanoparticle-decorated ZnS nanorod sensors to different ethanol concentrations at 300°C.

Fig. 7. (a) Response times and (b) recovery times of the pristine and $Cr_2O_3$ nanoparticle-decorated ZnS nanorod sensors towards different ethanol concentrations at 300°C.

Fig. 8. Responses of the pristine and $Cr_2O_3$ nanoparticle-decorated ZnS nanorod sensors to different VOC gases.

Fig. 9. (a) Energy band diagram of ZnS-$Cr_2O_3$ system before and after contact. (b) Schematic of the coross-section of the decorated nanorod and the energy band diagram of ZnS-$Cr_2O_3$ system showing the depletion layer width and potential barrier height.

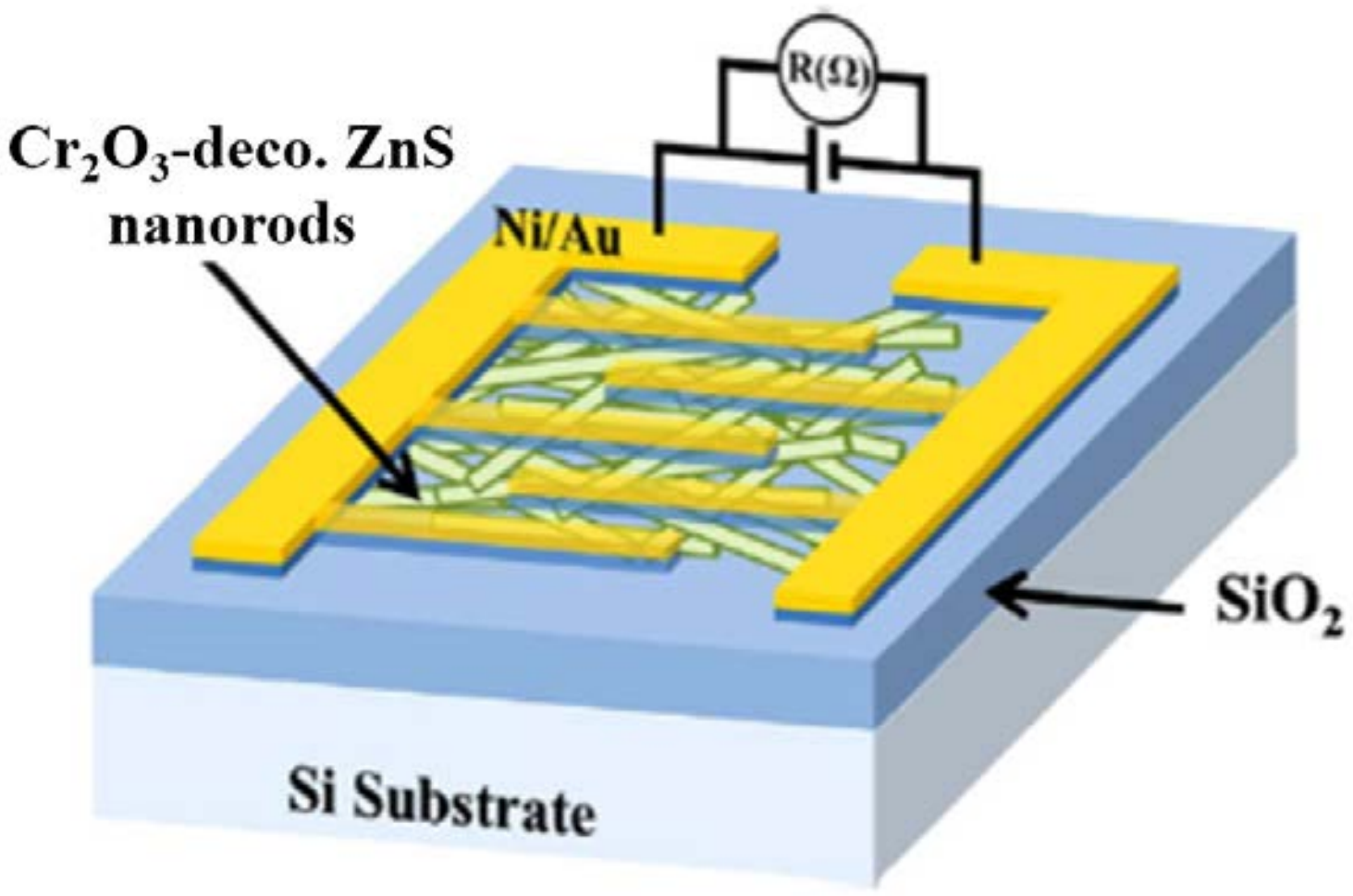

Fig. 1

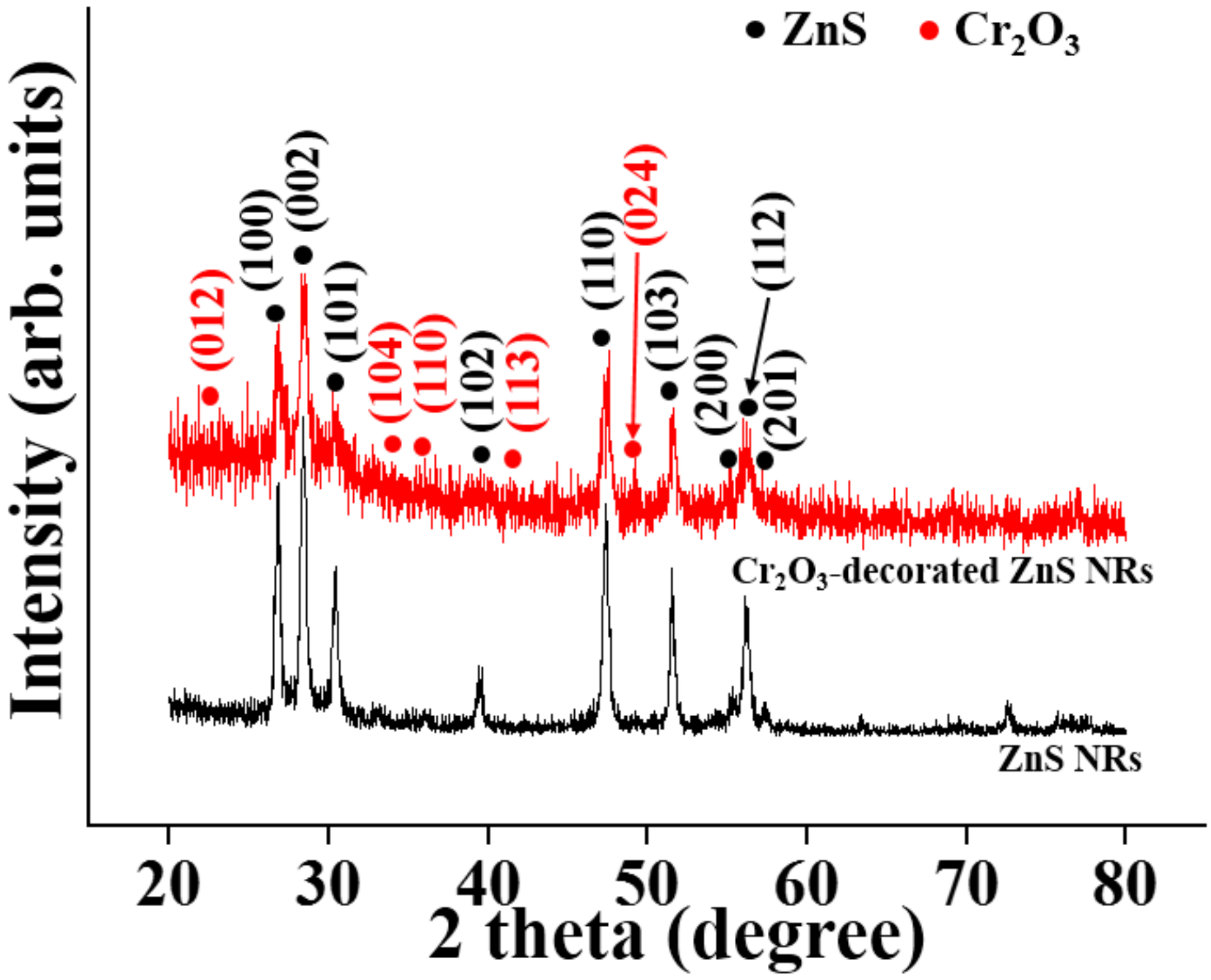

Fig. 2

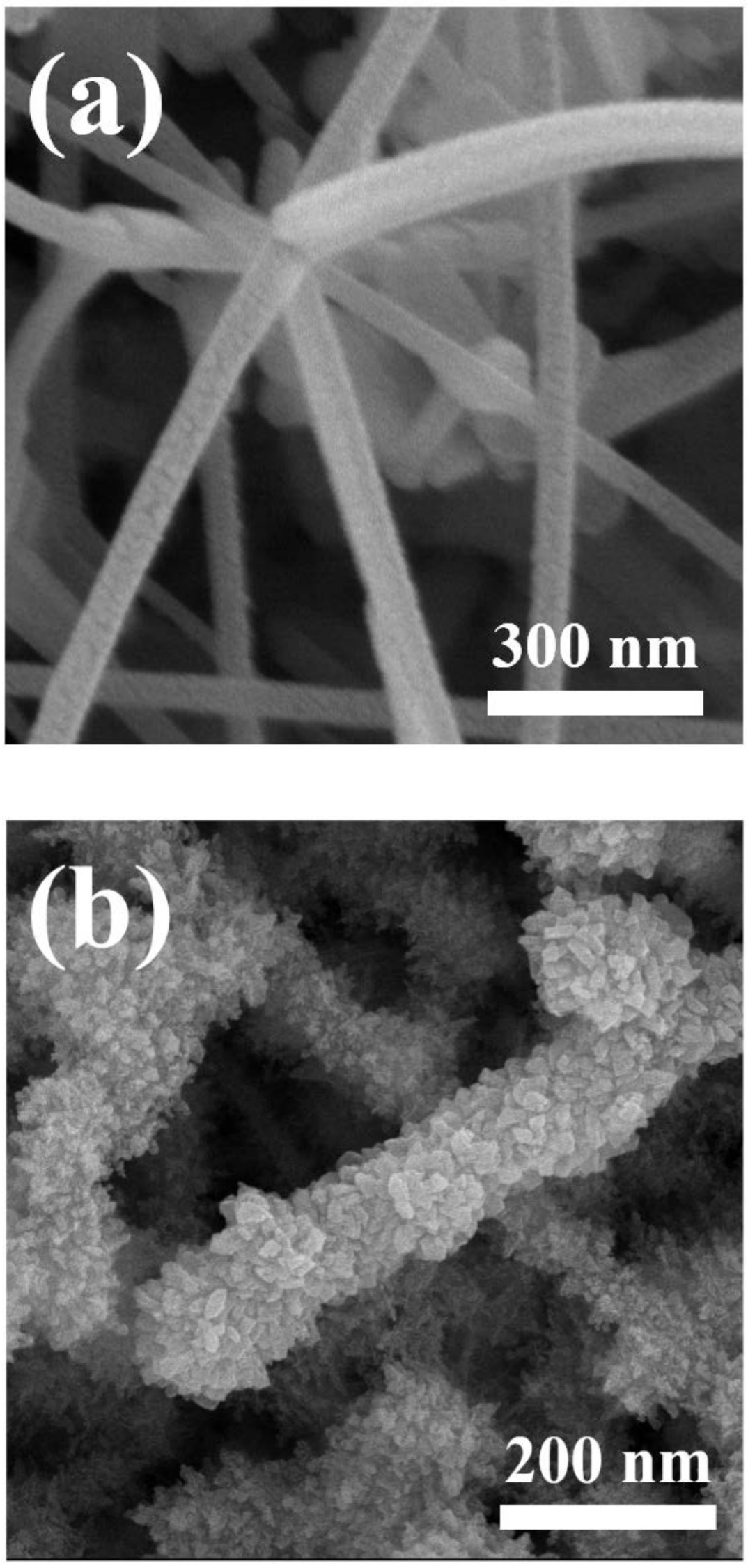

Fig. 3

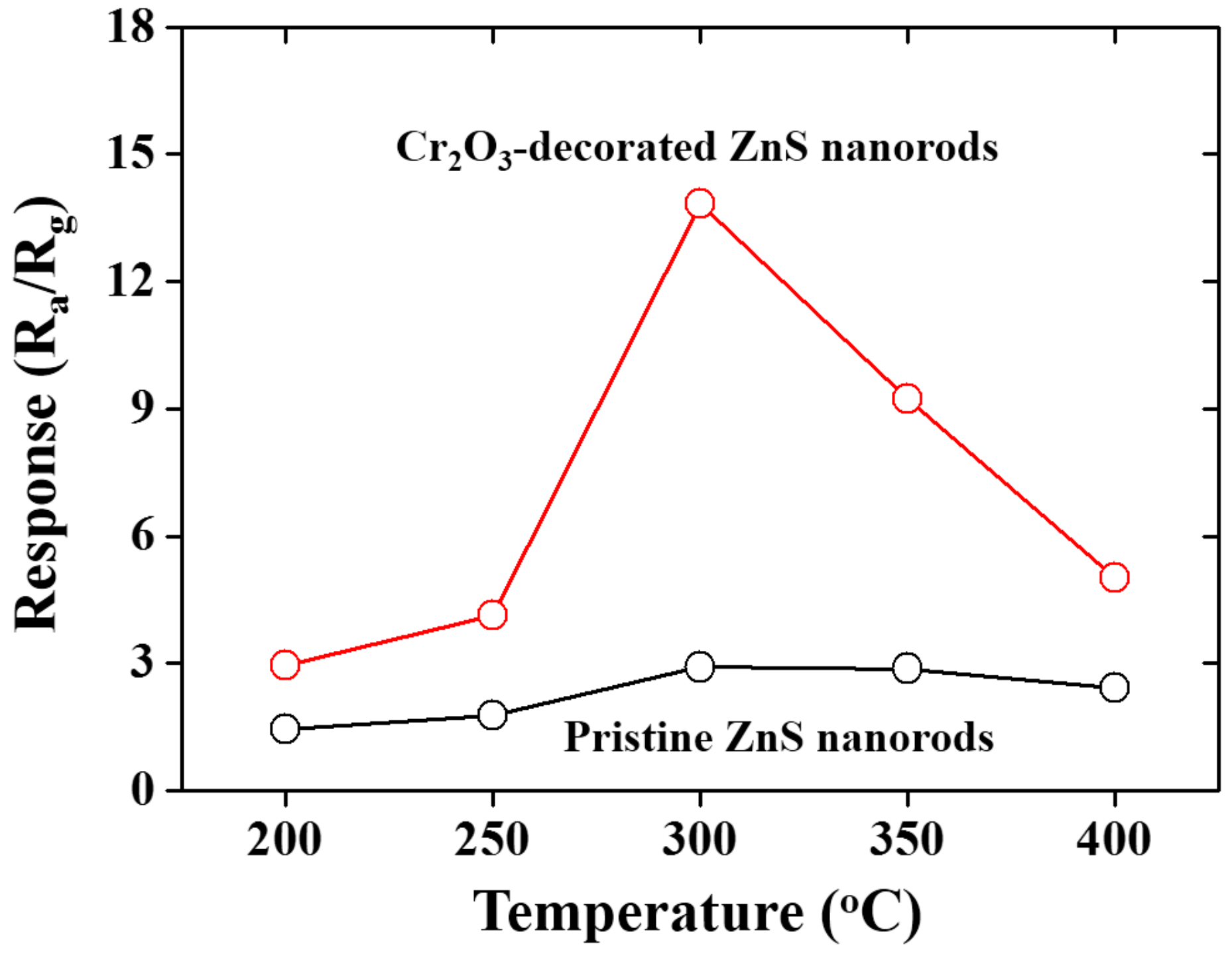

Fig. 4

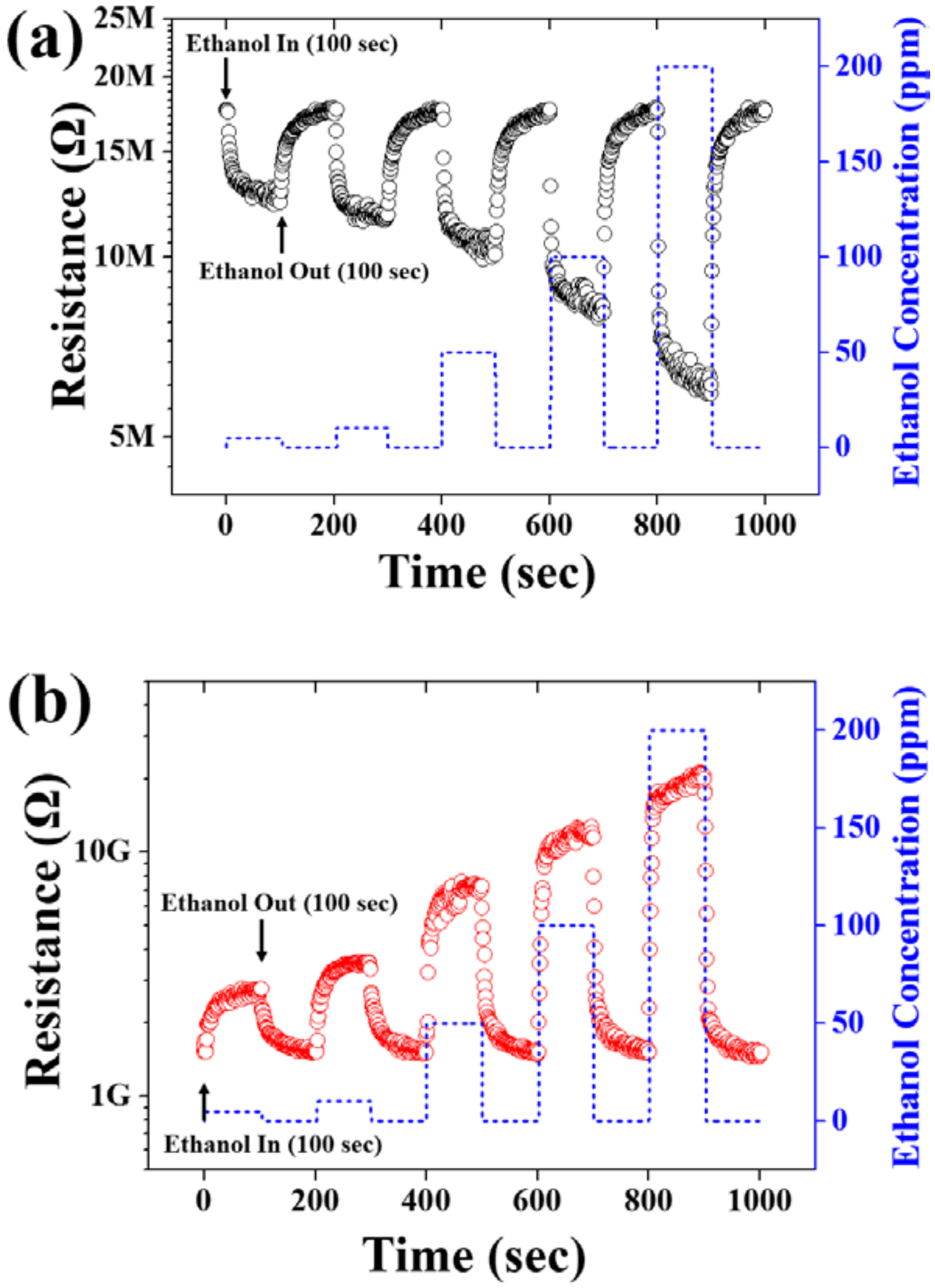

Fig. 5

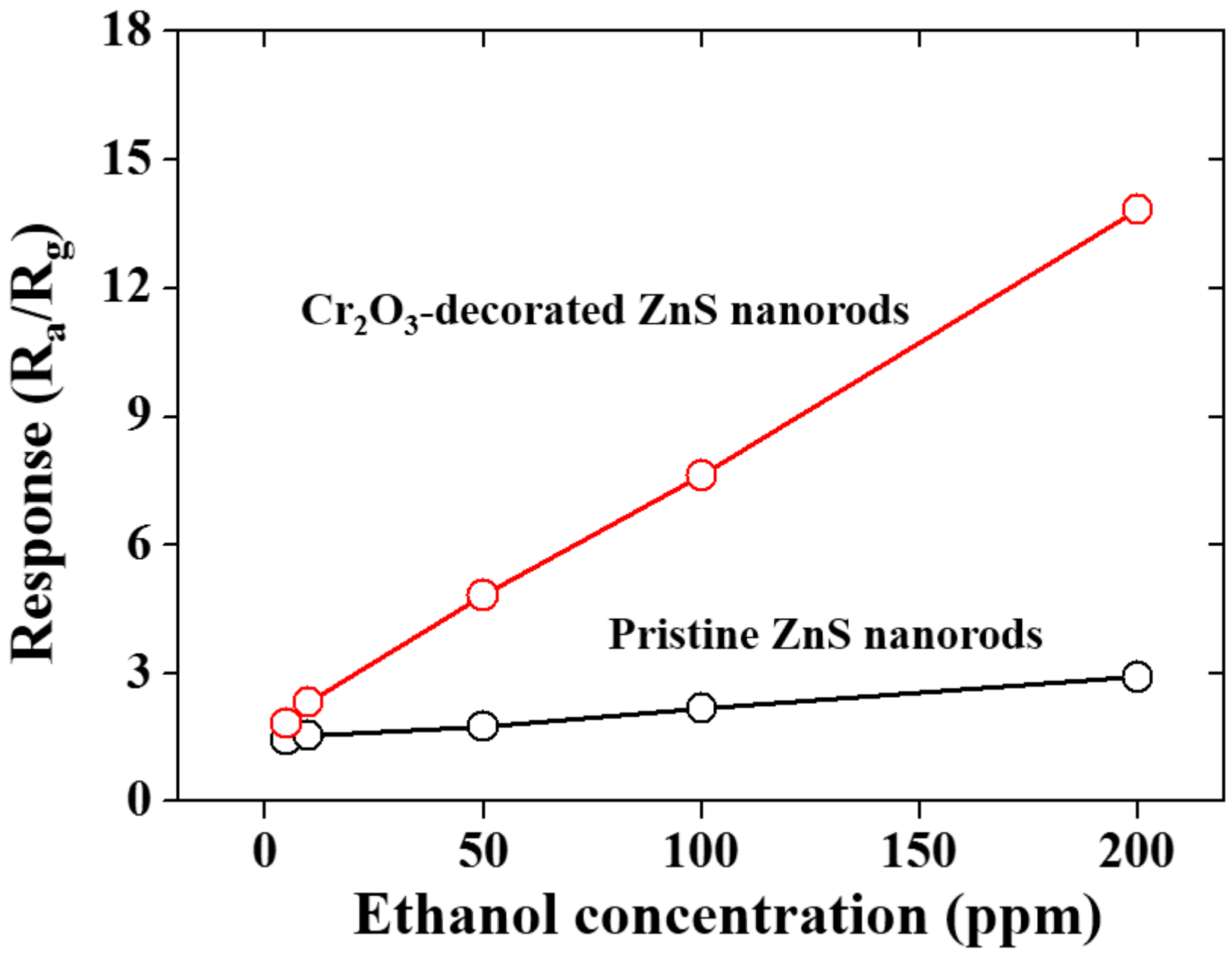

Fig. 6

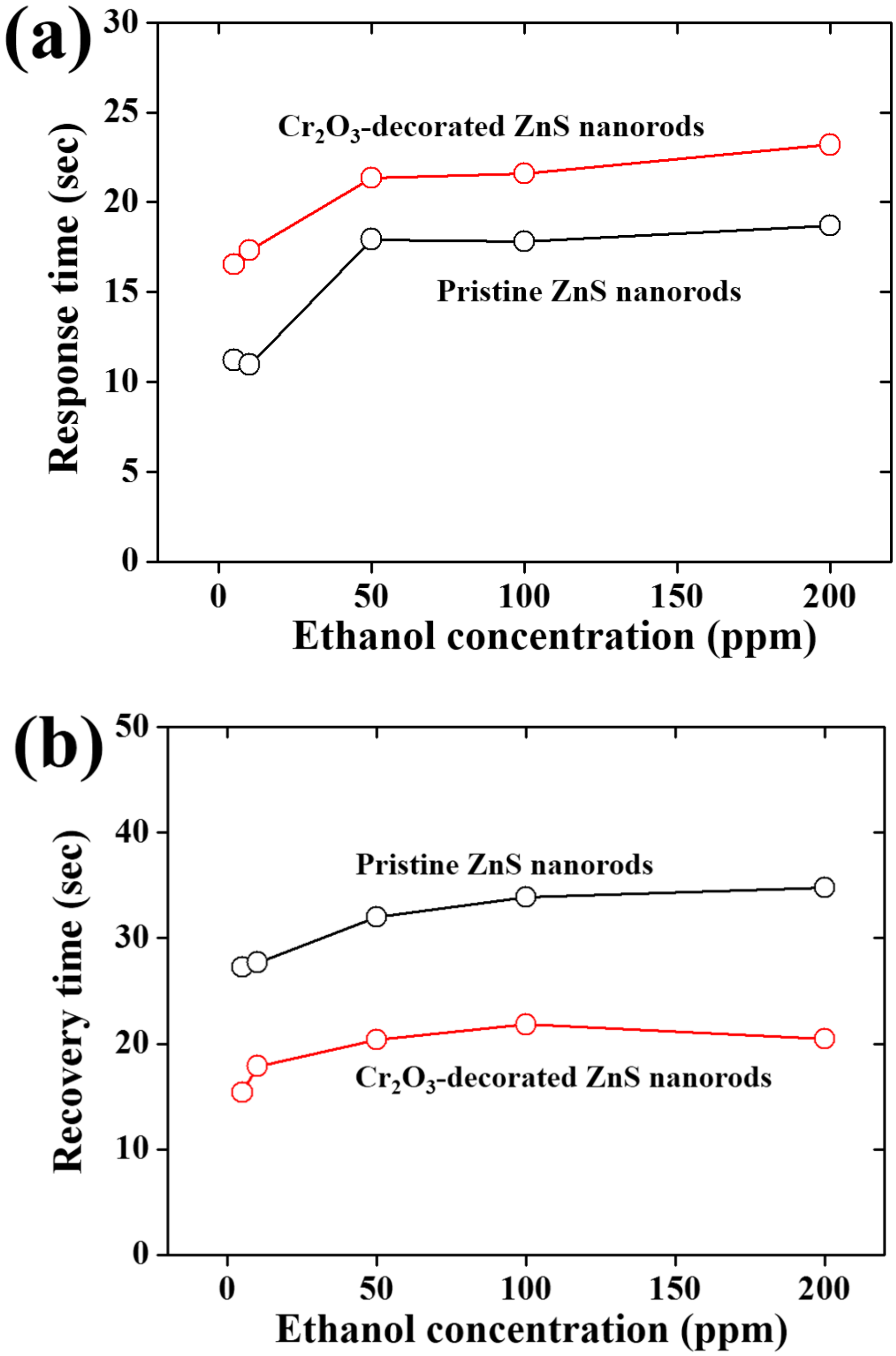

Fig. 7

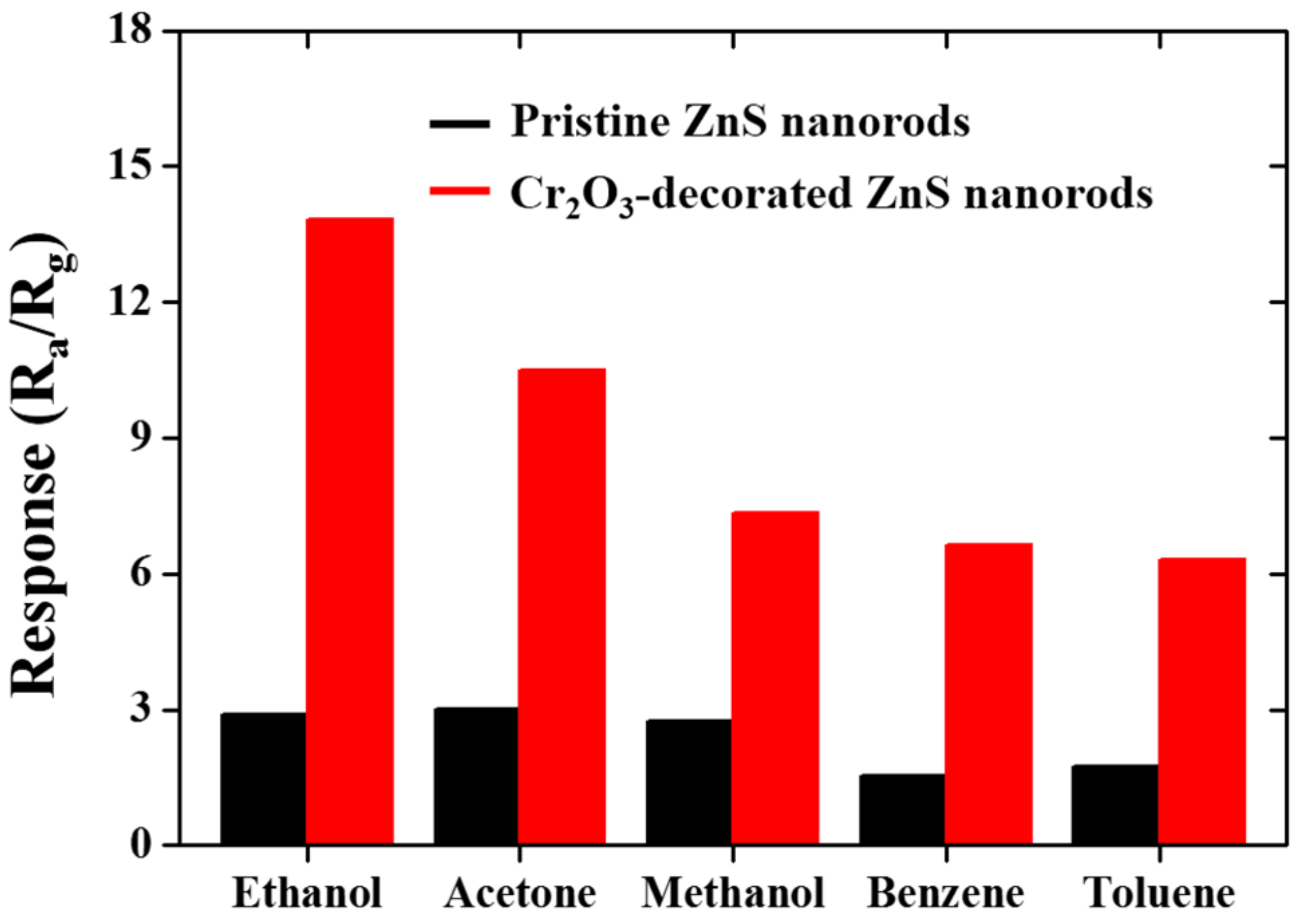

Fig. 8

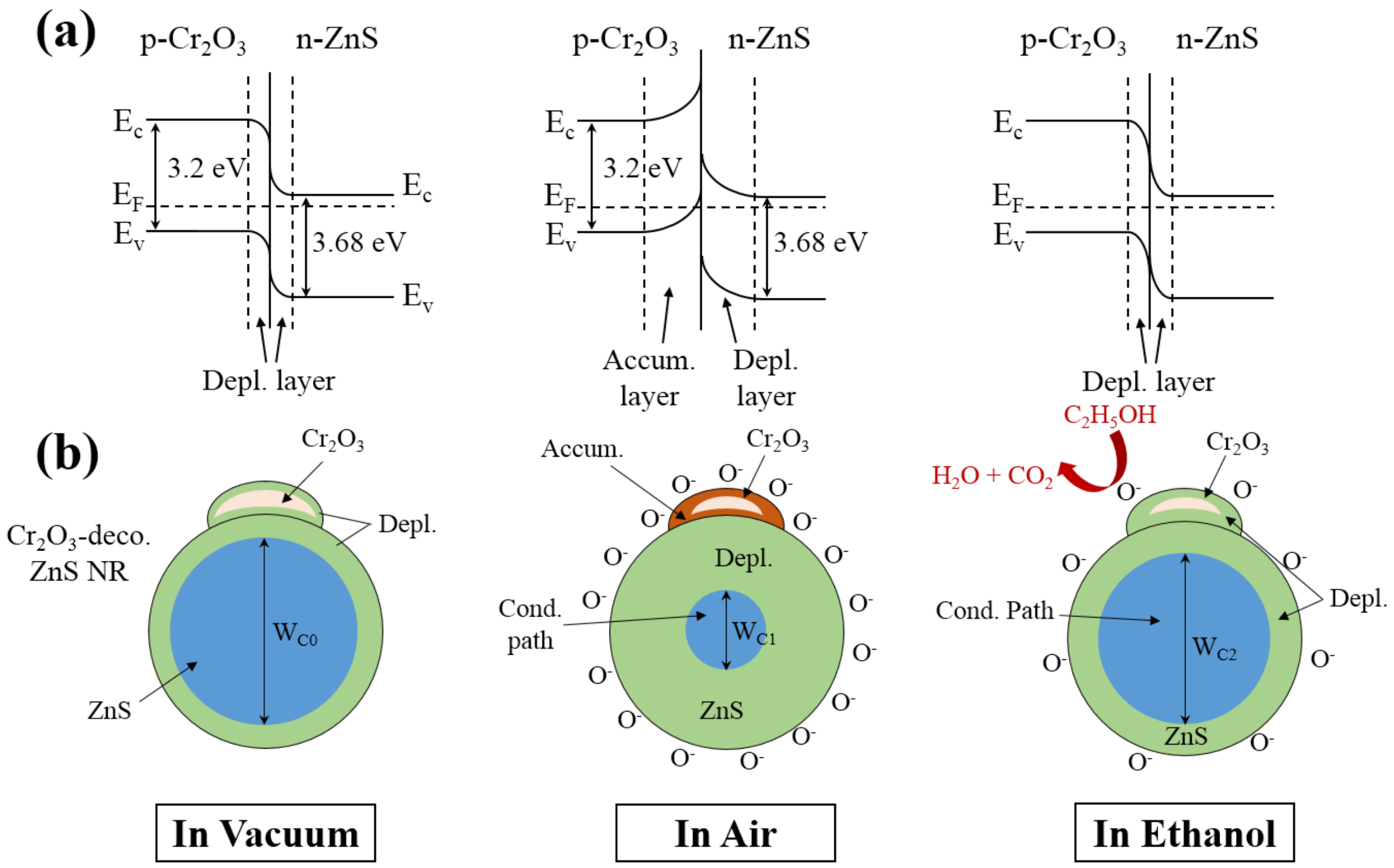

Fig. 9